# THE INTEGRATED MODEL OF cAMP-DEPENDENT DNA EXPRESSION REVEALS AN INVERSE RELATIONSHIP BETWEEN CANCER AND NEURODEGENERATION

Alfred Bennun, Ph.D.
Full Professor-Emeritus
Rutgers University, USA.

## Abstract

The model for cAMP-dependent synaptic plasticity relates the characterization of a noradrenaline-stimulated adenylyl cyclase with DNA unzipping. Specific proteins condition cAMP insertion in specific sites of the DNA structure to direct cellular and synaptic differentiation in brain tissues, also providing coding. Metabolic-dependent ATP binding of $Mg^{2+}$, could control feedback by inactivating AC dependent formation of cAMP. The level of cAMP and cGMP, which could be assayed in red cells and cerebrospinal fluid, allows a clinical lab diagnostic improvement. Also, provides a relationship of best fitting to cAMP control by binding to DNA. The cAMP level allows the prediction of an inverse relationship between neurodegeneration and cancer. The latter, could be characterized by uncontrolled proliferation, whereas metabolic dominance by stress over a long period of time, may deplete cerebral cAMP.

## Introduction

The study of noradrenaline (NA)-stimulated adenylyl cyclase (AC) [EC 4.6.1.1] of rat brain hypothalamus, cortex and corpus striatum [1], indicated a requirement for an excess of $Mg^{2+}$ over substrate MgATP. $Mg^{2+}$-ion at far greater concentrations resulted in basal activation curve of basal AC, but at considerable greater $Mg^{2+}$ excess over substrate.

DNA by heating at about 65°C results in uncoiling of its helix structure separating its two strands structure. The AC product, cAMP in conjunction with $Me^{2+}$ ($Mg^{2+}$, $Zn^{2+}$, $Mn^{2+}$) has a similar capacity for opening the structure of DNA [2]. Accordingly the tendency to form DNA-$Me^{2+}$-cAMP complexes uncoils a DNA segment, forming with the now separated strands a triple-stranded structure. Characterization as an epigenetic mechanism results from the ATP capacity to chelate the ligand metal, releasing cAMP from the complex.

This one allows to predict mechanisms for a role of cAMP on DNA unzipping and transcription. The red cell without mitochondria and DNA

controls the Pasteur Effect. This one shows that an $O_2$-dependent increase of citric cycle activity at the mitochondrial inhibits glycolysis at the cytoplasmic level.

The cAMP-ERK1/2-Bad signaling pathways by toxic levels of 6-OHDA (6-hydroxydopamine) a role in dopamine neuronal death of animal models for Parkinson's disease [3].

Uncontrolled cell division in cancer, circumvent apoptotic pathways, the patients show lower risk for neurodegenerative diseases, involving neuronal apoptosis, like Parkinson and Alzheimer [4]. This reverse relationship suggests sharing of a similar mechanism with different outcomes.

The reverse metabolic relationship characterizes cancer in many tissues. The $Mg^{2+}$ chelating capacity of ATP modulates the $Mg^{2+}$-dependent activity of adrenaline-stimulated AC [5] [6].

The nucleotide cAMP and cGMP bind DNA in a rather non-reversible and stable coding except in the presence of a stronger, chelating agent like $ATP^{4-}$.

This provides a metabolic connection for the cAMP level control of DNA expression, which at the brain could individualize neurons implicating the receptors proteins. Thus, allowing synaptic plasticity to function as memory networks.

## Results

Similarly, cAMP has been reported involved in CREB (cAMP response element-binding), protein that is involved as factor transcription and which is a part of memory function for many laboratories. This finding resulted in the development of behavior-cAMP linked models [2] [7] [8] [9] [10] [11] [12].

Kosmotropic (order-making) versus chaotropic (order-breaking) co-solvents influence equilibrium in aqueous solutions of proteins, acting through the structure and dynamics of the hydration water surrounding a protein [13]. The double hydration shell of $Mg^{2+}/Zn^{2+}/Me^{2+}$ allows conformational changes to enzymes and proteins, like hemoglobin. The turnover depends of an $Mg^{2+}$ binding to a protein requires losses on the hydration shell to form a chelate. This may have the tendency to decrease its hydration water and increasing the effective charge *nascent* ions like $Mg^{2+}$ could also explain $Zn^{2+}/Mn^{2+}$-fingers. Lately other groups provided a more fashionable denomination as $Zn^{2+}$ fingers, the chemistry is the same [14] [15] [16]. A divalent metal when attracted by more than one negative group constitutes a chelated state.

$Mg^{2+}$ when released by proteins like oxyHb turnover into deoxyHb appears in *nascent* state seeking water from the hydration shell of other ions like $Na^+$ and $K^+$.

These tendencies adjust the size of the hydration shells of $Na^+$ and $K^+$ ions for fitting into the gate-dynamics of the Na+-ion-pump.

The chaotropic strength could capture $H_2O$ from the kosmotropic $[K(H_2O)_6]^+$ to generate tetra-hydrated $[K(H_2O)_4]^+$ and tri-hydrated complexes $[K(H_2O)_3]^+$.

OxyHb regenerative effect on tissues at lower pH appears to depend on the release by red cells of $O_2$, $Mg^{2+}$ and cAMP/cGMP. The latter ones provide the DNA unzipping effect involved in transcription to generate the require polypeptides and proteins. In brain this coding effect creates memory at the level of individual neurons which function in neuronal networks [3].

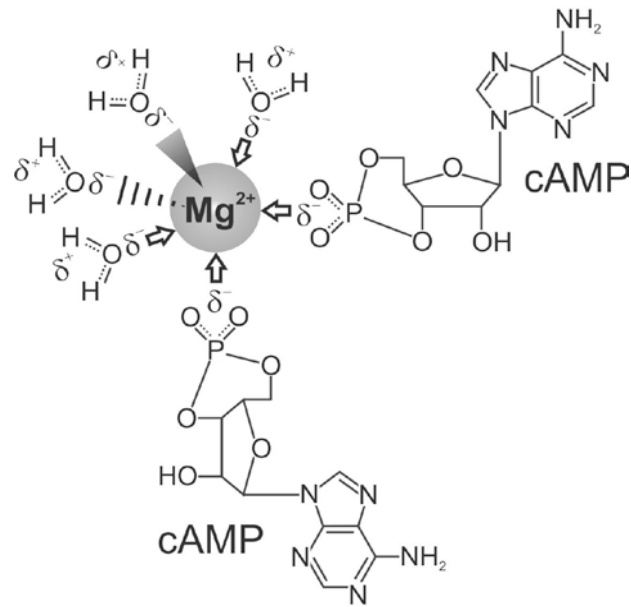

Figure 1. An $Mg^{2+}$-cAMP product, previously chelated by a protein R-groups, is released as an incomplete hydration shell. These promote an increase in the effective charge of the ion meriting the *nascent* denomination to $Mg^{2+}$. A following step could be the electrophilic attack on DNA by this *nascent* hydrated $Mg^{2+}$-cAMP enhanced simultaneously by coupling losses on the hydration shell.

The insertion of $Mg^{2+}$-cAMP opens the helix of the double-stranded DNA, to form a triple-stranded structure stabilized by pointing the purines and pyrimidines to the surface as showed in figure 3. This epigenetic mechanism provides coding for neuronal receptors.

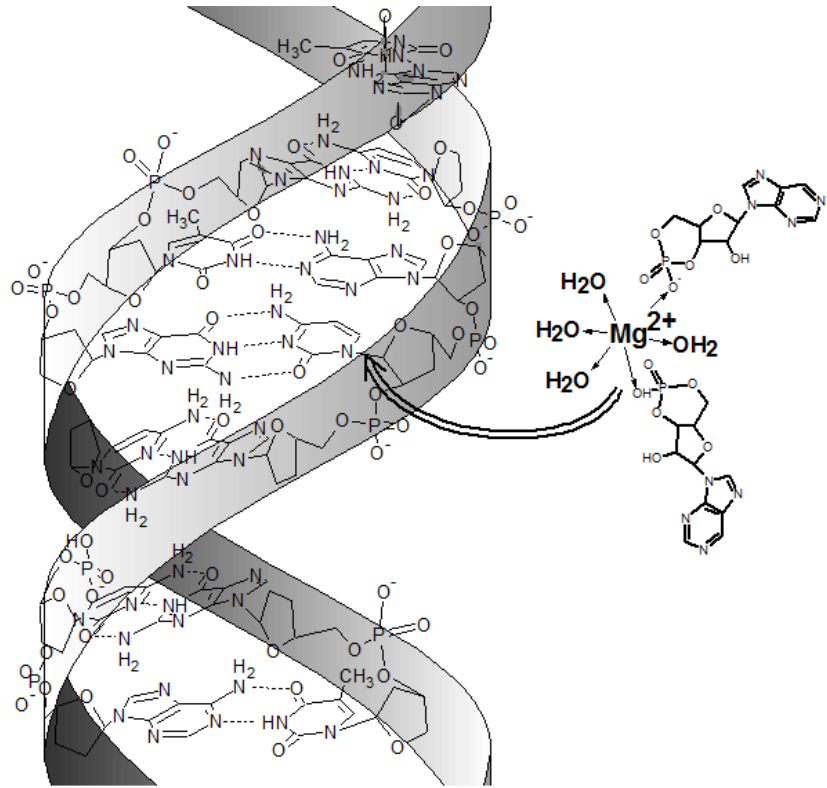

Figure 2. DNA-Mg-cAMP. The structure fitting function allows for DNA unzipping. cAMP/cGMP by binding to $Mg^{2+}$ ion interacts with the negatively charged phosphate groups of a segment of DNA, wherein the base sequence of the two strands has a binary rotational symmetry about a center (palindromic 5'-TGACGTCA-3'). This displays the mechanism of transcriptional activation function called CREs (cAMP-responsive enhancers).

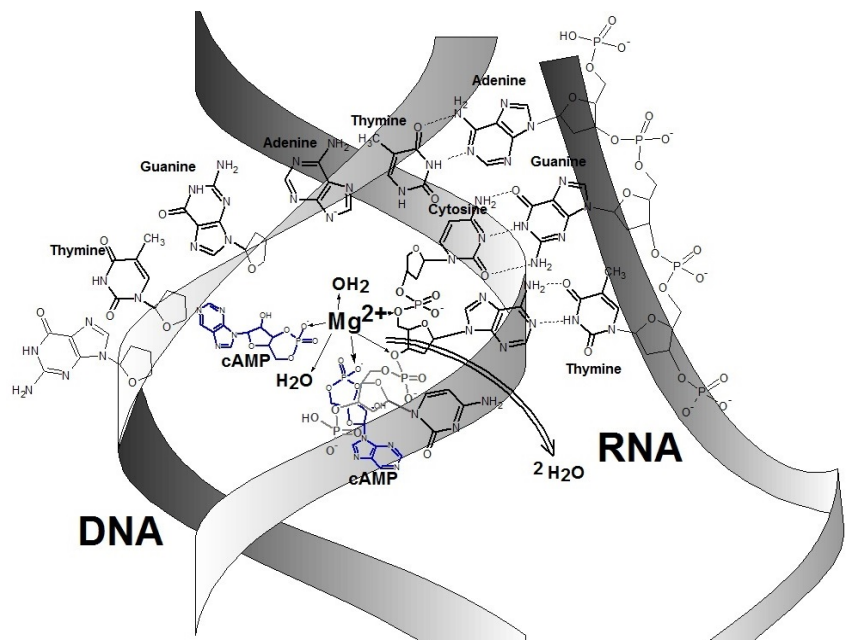

Figure 3. Crystallographic data has been used to find the fitting structure capable of playing the role of the CREB B-ZIP protein domain. In which a hexahydrated $Mg^{2+}$ ion binds with additional cAMP for maintaining open the double-stranded DNA containing the consensus CRE sequence (5'-TGACGTCA-3').

This structure is reversible by the releasing of the ion $Mg^{2+}$ by a negative charged molecule with great affinity for $Mg^{2+}$ like $ATP^{4-}$. This epigenetic model could serve to turn-on/off the DNA coding for the transcriptase reading. It could be applied to obtain a model of memory based in the natal language encryption readable on DNA.

During growing some of the first receptors expressed could be natal language and emotional experience. This could be used to assimilate receptors, within the first experiences required.

The learned meaning could be adapted to emotional experience, for example: it would be required to the transcriptase to express the oxytocin (bonded to the membrane by $Mg^{2+}$) sequence for a primary event like forming the receptors required for bonding between mother and recently born child.

Other emotions like reward could be expressed by the sequence of the dopamine receptor. The CREB B-ZIP protein domain could respond to specific proteins which indicate a locus in DNA for the receptor to be expressed. Learning of the natal language is a long process accelerated when reaching encoding of the neuronal network.

However, this encoding at specific neurons could direct selecting the neuronal network to be turn-on.

cAMP transport in the absence of DNA and AC activity explains the role of red cells in the tissue regeneration.

When $O_2$ is released by Hb its affinity for $Mg^{2+}$ is decreased, this became available for AC activation changing cellular cAMP levels. Uncontrolled process could trigger excessive cells duplication and cancer.

A negative charged $ATP^{4-}$ or similar compound can turn the process reversible by subtracting the divalent metals.

cAMP DNA insertion could be adapted in expression of specific polypeptide for each receptor.

## Conclusion

The search for models of the fitting structure between DNA and cAMP started by evaluate capability of $Mg^{2+}$ and cAMP to modify control of DNA expression.

A unified mechanism for cAMP controlled DNA-reading shows a non-covalent hydration modifying activation of the coded DNA expression by the fitting-in of cAMP on DNA.

cAMP and cGMP studies on red cells revealed interactions [17] [18] [19]. The absence of DNA indicated that should be transported to regions were DNA expression was required.

Protein and $H_2O$ are involved in energy expenditure during activation of protein synthesis turnover. In a protein the passage from its active to inactive states relates to the protein loss of H-bonds by a metabolic-dependent mechanism.

The release of water and metal by excess of ATP results on MgATP and the disappearance of free $Mg^{2+}$, which could inactivate AC and cAMP production [1] [2].

Attracting these groups resulted in the reorientation of the hydrophobic R-groups attracting between themselves.

The associated water through H-bonds leads to inactive architecture. In this context the H-bond loss from either hydrated state of proteins or DNA depends for turnover on entropy increase in the surrounding water [20]. This thermodynamics could characterize the homeostatic balance resulting from exhaled respiration saturated by vapor of water, eliminating entropy. The latter, do not contains H-bonds [21].

Assayed of cAMP and cGMP (Cyclic guanosine monophosphate) in red cells and cerebrospinal fluid (CSF) could be useful to obtain calibration tests, which physicians could relate to patients pathologic diagnosis.

Stress related diseases that through the fight-or-flight response mobilize metabolic reserves have a high consume phase that follows as an exhaustion phase. Oxytocin and serotonin adapt for modulated responses [22].

The testing of the cAMP and cGMP level should became a common practice in Hospitals because probably could contribute to early discover of neurodegenerative disorders. Subsequently, could allow for characterization of a disease progress and the probable outcome tendency.

Clinical could be expected that the assay of cAMP and cGMP in final phases of neurodegenerative diseases will show low concentration of this nucleotide. When as in Alzheimer in the brain appears empty areas, this could be related to the lack of regenerative protein capacity, because lack of cAMP turns inactive DNA. In Cancer development metabolic events leading to cell proliferation, may result from accumulation of cAMP and cGMP [23], showing uncontrolled transcription.

It could be expected that the assay of cAMP and cGMP in red cells and CSF could allow detecting in brain neurodegenerative diseases.

These characterizing these diseases by nucleotide value when well above the norm. Samples values correlating in a curve could allow the reading of results, in the detecting correlation curves. These results could characterize dysfunctions and the specific value when in deviation.

Similar effect could lead to the characterization of specific cancer [24], which could be probed by reagents testing for cAMP-cGMP in red cells and for neurodegeneration in CSF.

**References**

[1] Brydon-Golz, S. and Bennun, A., Postsynthetic stabilized modification of adenylate cyclase by metabolites, Biochemical Society Transactions, 3, (1975), 721-724.
[2] Brydon-Golz, S., Ohanian, H. and Bennun, A., Effects of noradrenaline on the activation and the stability of brain adenylate cyclase, *Biochem. J*., 166, (1977), 473-483.
[3] Bennun A. Book: Molecular Aspects of the Psychosomatic-Metabolic Axis and stress. Series: Neurology - Laboratory and Clinical Research Developments. Editorial: Nova Science Publishers, 2015. *ISBN: 978-1-63463-912-5*.
[4] The New York Academy of Sciences. Learning from Cancer to Advance Drug Development for Neurodegeneration. Academy eBriefings. 2015. Available at: www.nyas.org/BDDGCancer-eB
[4] Silva, A.J., Kogan, J. H., Frankland, P.W. & Kida, S. (1998) CREB and memory. *Annual Review of Neuroscience* 21, 127-148.

[5] Harris, R. and Bennun, A., Hormonal control of fat cells adenylate cyclase, *Molecular & Cellular Biochemistry*, 13, (1976), No. 3, 141-146.

[6] Harris, R.H., Cruz, R. and Bennun, A., The effect of hormones on metal and metal-ATP interactions with fat cell adenylate cyclase, Biosystems, 11, (1979), 29-46.

[7] Silva, A.J., Kogan, J. H., Frankland, P.W. & Kida, S. (1998) CREB and memory. Annual Review of Neuroscience 21, 127-148.

[8] Davis, H. P. and Squire, L. R. (1984) Protein synthesis and memory: a review. Psychol. Bull. 96, 518–559.

[9] Mayr, B. and Montminy, M. (2001) Transcriptional regulation by the phosphorylation-dependent factor CREB. Nat. Rev. Mol. Cell Biol. 2(8), 599-609.

[10] Kandel, E.R. (2012) The molecular biology of memory: cAMP, PKA, CRE, CREB-1, CREB-2, and CPEB. *Mol Brain* 5(1), 14.

[11] Kandel, E. R. (2001) The molecular biology of memory storage: a dialogue between genes and synapses. *Science* 294, 1030–1038.

[12] Bennun, A. and Blum, J.J. (1966) Properties of the induced acid phosphatase and of the constitutive acid phosphatase of Euglena. *Biochimica et Biophysica Acta*, 128, 106-123.

[13] Russo, D. (2008) The impact of kosmotropes and chaotropes on bulk and hydration shell water dynamics in a model peptide solution. *Chemical Physics* 345, 200–211.

[14] Bennun, A., A coupling mechanism to inter-relate regulatory with haem-haem interactions of haemoglobin, *Biomed. Biochim. Acta*, 46, (1987), No. 2/3, 314-319.

[15] Emma, J.E., Cervoni, P., Sulner, J.W. and Bennun, A., KCI-stimulated renin release, *Annals of the New York Academy of Sciences*, 463, (1986), 281-283.

[16] Bennun, A., Seidler, N. and De Bari, V.A., Divalent metals in the regulation of hemoglobin affinity for oxygen, Annals of the New York Academy of Sciences, 463, (1986), 76-79.

[17] DeBari, V. A. and Bennun, A. (1982). Cyclic GMP in the human erythrocyte. Intracellular levels and transport in normal subjects and chronic hemodialysis patients. Clin. Biochem., 15, 219-221.

[18] Novembre, P.; Nicotra, J.; DeBari, V. A.; Needle, M. A. and Bennun, A. (1984). Erythrocyte transport of cyclic nucleotides. Annals of the New York Academy of Sciences, 435, 190-194.

[19] DeBari, V. A.; Novak, N. A. and Bennun, A. (1984). Cyclic Nucleotide Metabolism in the Human Erythrocyte. Clin. Physiol. Biochem., 2, 227-238.

[20] Bennun A. cAMP-Me2+-DNA Complex on Gene Induction and Signaling for Coupling the Environment Stimulus to Produce Variety and

its Impact on Evolution. *Advances in Medicine and Biology*. Volume 98, Chapter10, Number xx, Pages xx-xx (2016).

[21] Bennun A. The coupling of thermodynamics with the organizational water-protein intra-dynamics driven by the H-bonds dissipative potential of water cluster. Nova Publishers. Submitted 07/04/2016.

[22] Bennun A. The Noradrenaline-Adrenaline-Axis of the Fight-or-Flight Exhibits Oxytocin and Serotonin Adaptive Responses. International Journal of Medical and Biological Frontiers. Volume 21, Issue 4, pages: 387-408 (2015).

[23] Fajardo, A. M.; Piazza, G. A.; Tinsley, H. N. (2014) The role of cyclic nucleotide signaling pathways in cancer: targets for prevention and treatment. *Cancers (Basel)*. 6(1):436-58.

[24] Miller, C. A.; Sweatt J. D. (2008) Covalent Modification of DNA Regulates Memory Formation. *Neuron*. 59(6), 1051.